\documentclass[twocolumn,prl,nobalancelastpage,aps,10pt]{revtex4-1}
\usepackage{graphicx,bm,times}

\usepackage{natbib} \usepackage{miller}
\begin{document}

\title{The Role of Crystal Orientation in the Dissolution of UO$_2$ Thin Films}

\author{S. Rennie$^{\dag}$, E. Lawrence Bright$^{\dag}$, J. E. Sutcliffe$^{\dag}$, J. E. Darnbrough$^{\dag}$, J. Rawle$^{\ddag}$, C. Nicklin$^{\ddag}$, G. H. Lander$^{\S}$ and R. Springell$^{\dag}$.}

\affiliation{$^{\dag}$ University of Bristol Interface Analysis Centre, HH Wills Physics Laboratory, Tyndall Avenue, Bristol, BS2 8BS UK.
\\
$^{\ddag}$ Diamond Light Source, Harwell Science and Innovation Campus, Fermi Ave, Didcot OX11 0DE.
\\
$^{\S}$ European Commission, Joint Research Centre (JRC), Directorate for Nuclear Safety and Security, Postfach 2340, D-76125 Karlsruhe, Germany.
}
\begin{abstract}
Epitaxial thin films have been utilised to investigate the radiolytic dissolution of uranium dioxide interfaces. Thin films of UO$_2$ deposited on single crystal yttria stabilised zirconia substrates have been exposed to water in the presence of a high flux, monochromatic, synchrotron x-ray source. In particular, this technique was applied to induce dissolution of three UO$_2$ thin films, grown along the principle UO$_2$ crystallographic orientations: (001), (110) and (111). Dissolution of each film was induced for 9 accumulative corrosion periods, totalling 270\,s, after which XRR spectra were recorded to observe the change in morphology of the films as a function of exposure time. While the (001) and (110) oriented films were found to corrode almost linearly and at comparable rates, the (111) film was found to be significantly more corrosion resistant, with no loss of UO$_2$ material being observed after the initial 90s corrosion period. These results distinctly show the effect of crystallographic orientation on the rate of x-ray induced UO$_2$ dissolution. This result may have important consequences for theoretical dissolution models, as it is evident that orientation dependence must be taken into consideration to obtain accurate predictions of the dissolution behaviour of UO$_2$.
  \end{abstract}
\date{\today}

\maketitle

\section{INTRODUCTION}
Understanding the corrosion behaviour of nuclear reactor fuels throughout the fuel cycle is key to assessing safety considerations \cite{He2012}. In particular, investigating the interaction of fuel in the presence of water is an important scenario that has been widely studied \cite{Shoesmith2000,Sunder1997,Matzke1992,Shoesmith2007}, given it's potential to occur both within the reactor under accident conditions and during the storage of spent nuclear fuel.

The most commonly used nuclear fuel, uranium dioxide, has a low solubility in water, given that the uranium exists in the (IV) oxidation state \cite{Shoesmith1992}. However, in the presence of radiation fields, water undergoes radiolysis to produce a wide range of highly oxidising species, including: H$_2$O$_2$, OH$^{\bullet}$, O$_2^{\bullet-}$, HO$_2^{\bullet}$ and O$_2$ \cite{LeCaer2011, Spinks1990, Shoesmith2007}. In the presence of these species, UO$_2$ is readily converted to the uranyl ion (UO$_{2}^{2+}$) via the oxidation of U(IV) to U(VI), which has a solubility several orders of magnitude greater than UO$_2$ \cite{Shoesmith1992,Bailey1985}. This process can lead to the accelerated dissolution of the fuel matrix and potential release of radionuclides \cite{Shoesmith2007}.

Studying the dissolution of spent nuclear fuel in an aqueous environment is challenging, owing to the complex nature of both the fuel itself and the unique environments present throughout the fuel cycle. A large body of work has been conducted internationally to investigate the dissolution of UO$_2$ under a vast range of conditions \cite{Sunder2004,Shoesmith1991,Shoesmith1989,Sunder1991,Bailey1985,Clarens2005,Shoesmith2007,Ekeroth2006,Eriksen2012}, however, understanding the isolated effect of individual parameters is difficult, and single variable studies are required.  In this regard, thin films offer considerable potential, providing highly versatile, idealised samples that can be readily engineered to explore a wide range of material properties. Conducting nuclear material research on these systems presents the opportunity to begin with a model surface, and gradually introduce complexity such that a more fundamental understanding of fuel material properties can be obtained. This approach has particular advantages in supporting theoretical models, as such studies are typically performed on idealised surfaces, for which long-term simulations can be tested. In contrast with experiments conducted on complex bulk oxide materials or simulated fuels, our studies bridge the gap between theory and realistic conditions, with the aim of obtaining a more fundamental understanding of interactions at the fuel-water interface.

With recent developments in the growth of single crystal UO$_2$ thin films \cite{Strehle2012,Bao2013}, experimental studies on these idealised surfaces are now being performed \cite{Springell2015,Teterin2016,Popel2017}. In particular, we have utilised synchrotron radiation to induce dissolution at the UO$_2$\,/\.water interface. This approach \cite{Springell2015}, successfully induced radiolytic dissolution of UO$_2$, and enabled the change in the surface morphology of the film to be subsequently probed using synchrotron x-ray reflectivity (XRR) measurements. The work presented here, builds on our previous work to investigate the effect of crystallographic orientation on the dissolution of UO$_2$.

\section{EXPERIMENTAL DETAILS}
\subsection{Sample Growth}

\begin{figure*}
	\centering
	\includegraphics[height=0.22\textheight]{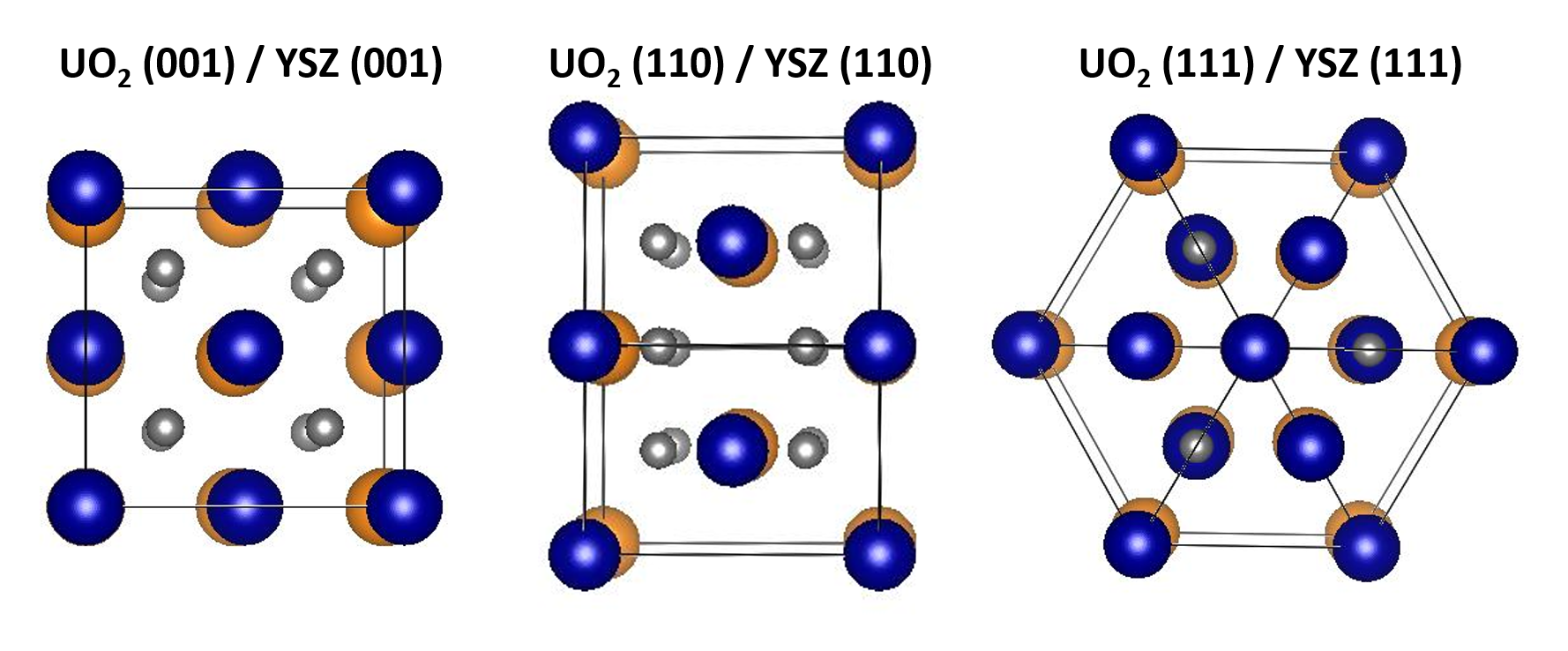}
	\caption{Deposition of the three principle UO$_2$ orientations (001), (110) and (111) was achieved through using (001), (110) and (111) oriented YSZ substrates. The lattice mismatch is the same in each case at 6.3 $\%$. Both the (001) and (111) surfaces are non-polar, whereas the (110) is polar \cite{Bottin2016}. Figures produced using the VESTA software \cite{Momma2011}.}
	\label{fig:ysz_sub}
\end{figure*}

Epitaxial UO$_2$ thin films were grown at the University of Bristol using DC magnetron sputtering. An argon plasma sustained using a gas pressure of P$_{Ar} =$ 7.2$\times$10$^{-3}$\,mbar was utilised to sputter atoms from a uranium metal target in an oxygen partial pressure of P$_{O2} =$ 2$\times$10$^{-5}$\,mbar, enabling UO$_2$ to be deposited. Under these conditions, UO$_2$ was sputtered at a rate of 1.6 \AA\, / s onto substrates held at a temperature of 1000$\,^{\circ}$C; where the thermal energy was provided to improve the crystallinity of the film.

To investigate the effect of crystallographic orientation on UO$_2$ dissolution, epitaxial UO$_2$ thin films were grown for each principal crystal orientation: (001), (110), and (111); where the orientation was determined through the choice of substrate. Given both UO$_2$ and yttria stabilized zirconia (YSZ) posses the cubic fluorite crystal structure with similar bulk lattice parameters of 5.47\,\AA\, and 5.15\,\AA\, respectively, a direct epitaxial match can be obtained for each principle orientation, Fig \ref{fig:ysz_sub} \cite{Strehle2012}. Atomically polished, 10 $\times$ 10 $\times$ 0.5\,mm, YSZ (001), (110) and (111) crystals obtained commercially from MTI corp., were therefore selected as substrates on to which epitaxial UO$_2$ thin films, of approximately 80\,\AA\, in thickness, were deposited.

\subsection{Radiolytically Induced Dissolution}
To simulate radiolytic dissolution of UO$_2$ fuel, epitaxial UO$_2$ thin films were exposed to a layer of pure (MilliQ) water in the presence of a synchrotron x-ray beam. Experiments were conducted on the I07 beamline at the Diamond Light Source, utilising a 17.116\,keV x-ray beam focussed in both the horizontal and vertical directions to give a beam size of 100\,$\mu$m\,$\times$\,100\,$\mu$m and a flux of 1x10$^{14}$\,m$^{-2}$s$^{-1}$ at the sample position. The beam energy was specifically selected to be 50\,eV below the \textit{U L$_3$} absorption edge in order to prevent excitations in the UO$_2$ surface. Slit conditions were chosen to produce a square beam profile, limiting the variation in x-ray intensity across the irradiated region.

\begin{figure}[h]
	\centering
	\includegraphics[height=0.2\textheight]{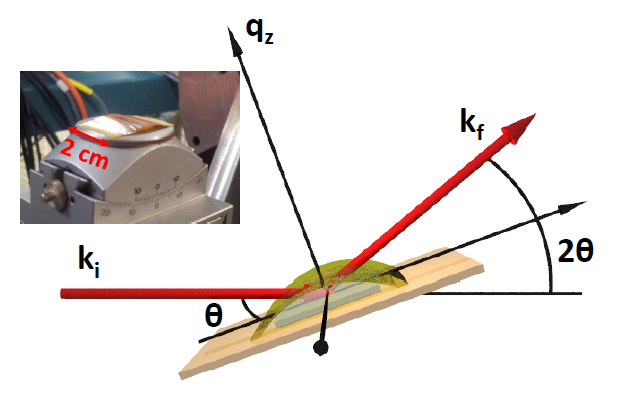}
	\caption{Schematic of the experimental set up used on beamline I07 at the Diamond Light Source to conduct x-ray induced radiolytic dissolution of epitaxial UO$_2$ thin films. The thin layer surface tension cell, additionally shown in the insert photograph, holds a fixed volume of water over the sample surface during the x-ray irradiation. To characterise the structure of the films XRR and XRD measurements were conducted in a specular geometry. Incident and exit wavevectors are labelled k$_i$ and k$_f$, at angle $\theta$ to the film surface, give rise to a the wavevector momentum transfer (q$_z$) in the specular direction \cite{Springell2015}.}
	\label{fig:thinfilmcell}
\end{figure}

To maintain a fixed volume of water over the sample surface, the film was held within a thin layer surface tension cell, as depicted in Fig. \ref{fig:thinfilmcell}. The cell comprises a domed Kapton film extended over the thin film sample, secured to a stainless steel stub using an adhesive tape. MilliQ water was introduced beneath the Kapton film, such that the entire UO$_2$ film surface was in contact with the MilliQ water solution. With the film entirely covered by a layer of water, an intense, monochromatic x-ray beam was directed at the sample surface, inducing water radiolysis and stimulating the production of the oxidation products required for the dissolution process to occur.

\begin{figure}
	\centering
	\includegraphics[height=0.7\textheight]{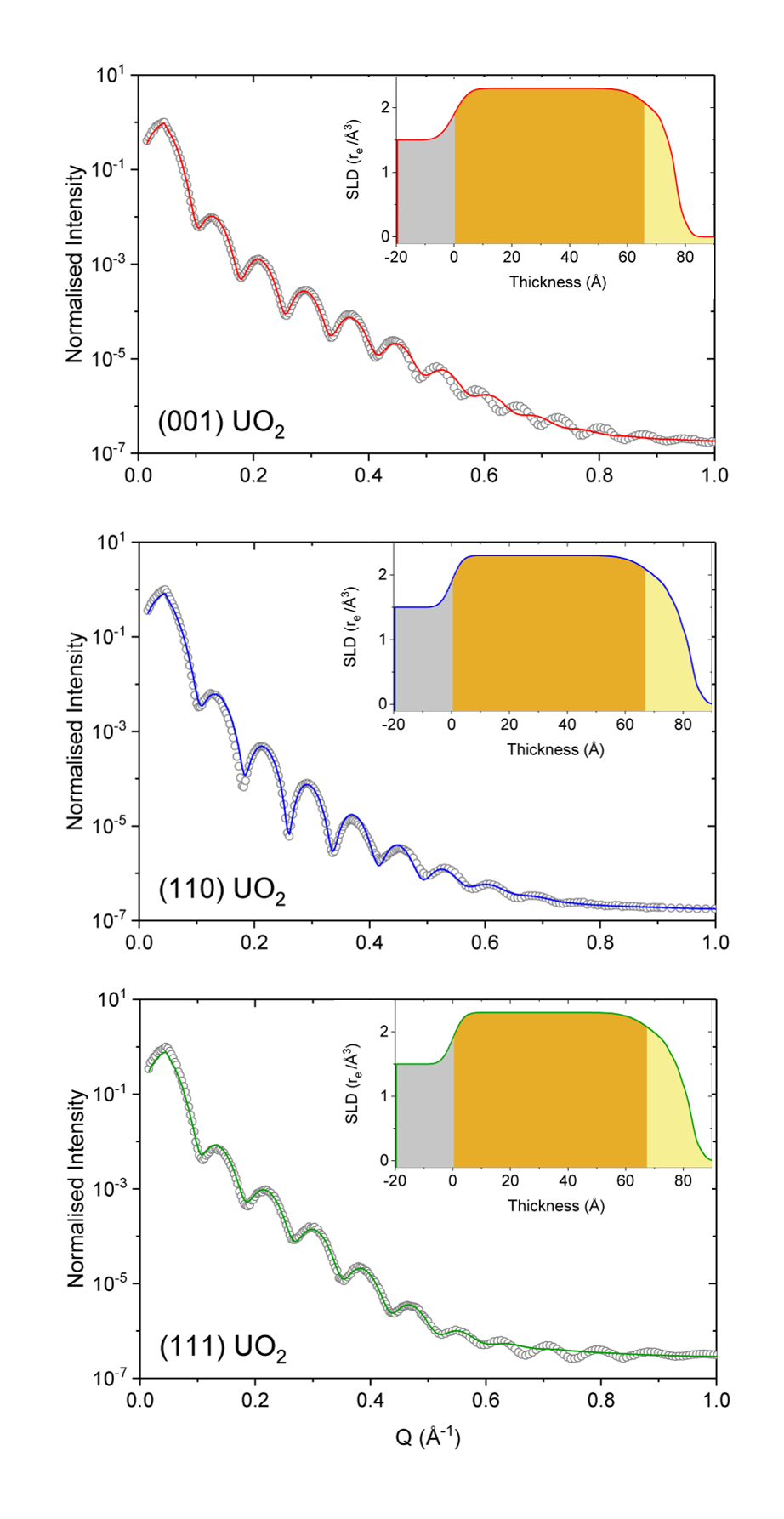}
	\caption{X-ray reflectivity spectra of the pristine (001), (110) and (111) oriented UO$_2$ thin films. The experimental data are represented by the open grey circles and the fitted calculation by the solid red, blue and green lines respectively. For each fitted reflectivity profile, the corresponding scattering length density (SLD) plot is shown as an inset. Here, 0 \AA\, represents the substrate / UO$_2$ interface and the fitted thickness parameters for the UO$_2$ and UO$_X$ layers are represented by the dark and light yellow areas, respectively, with the grey area representing the substrate.}
	\label{fig:pristinexrr111}
\end{figure}

In addition to providing the intense radiation fields required to induce water radiolysis, this technique additionally utilises the x-rays to monitor the change in morphology of the UO$_2$ surface.  This dual source-probe approach, utilises x-ray reflectivity (XRR) in a specular geometry. These surface sensitive techniques allow for changes in the electron density, surface roughness, interface structure, crystallinity and eventual dissolution of the UO$_2$ film to be probed as a function of exposure time. The ability to access such a wide range of information, enables detailed observation of the dissolution process as it unfolds at the surface of the UO$_2$ film. This level of detail cannot be observed using a bulk UO$_2$ sample; considering a typical energy range for a synchrotron single crystal diffraction beamline (10 - 20 keV) and the angles of incidence required to measure the associated reflections for UO$_2$, the penetration of the x-ray beam into the UO$_2$ is of the order of several microns. Therefore, any changes that occur on the surface of the sample, where the dissolution is concentrated, will be masked by the signal generated from the unchanging bulk. However, by conducting such experiments on a UO$_2$ thin film sample, with a thickness in the order of $\sim$\,100\,\AA, any changes occurring on the surface of the film will be evident. Implementing a thin film approach therefore provides unique sensitivity to any induced structural changes taking place at the UO$_2$ surface. This, coupled with the ability to utilise epitaxial thin films to isolate the effect of crystallographic orientation, makes this developed methodology particularly advantageous for conducting UO$_2$ dissolution studies.

\section{RESULTS}
\subsection{Sample Characterisation}

Prior to the dissolution experiment, XRR was used to determine the starting layer thickness and roughness for each sample. The recorded XRR spectra were fitted using the GenX computer program \cite{Bjorck2007}, which generates theoretically calculated reflectivity profiles using the Parratt recursion method of calculating transmitted and reflected wave fields \cite{Parratt1954}. The calculated profiles are fitted to the experimental data using a differential evolution algorithm; in order to avoid local minima, a problem that is regularly encountered when modelling x-ray reflectivity spectra. The fits presented here use the `simplex best 1 bin' fitting method and a figure of merit (FOM) of LogR1 \cite{Bjorck2007}.

\begin{figure}[h]
	\centering
	\includegraphics[height=0.7\textheight]{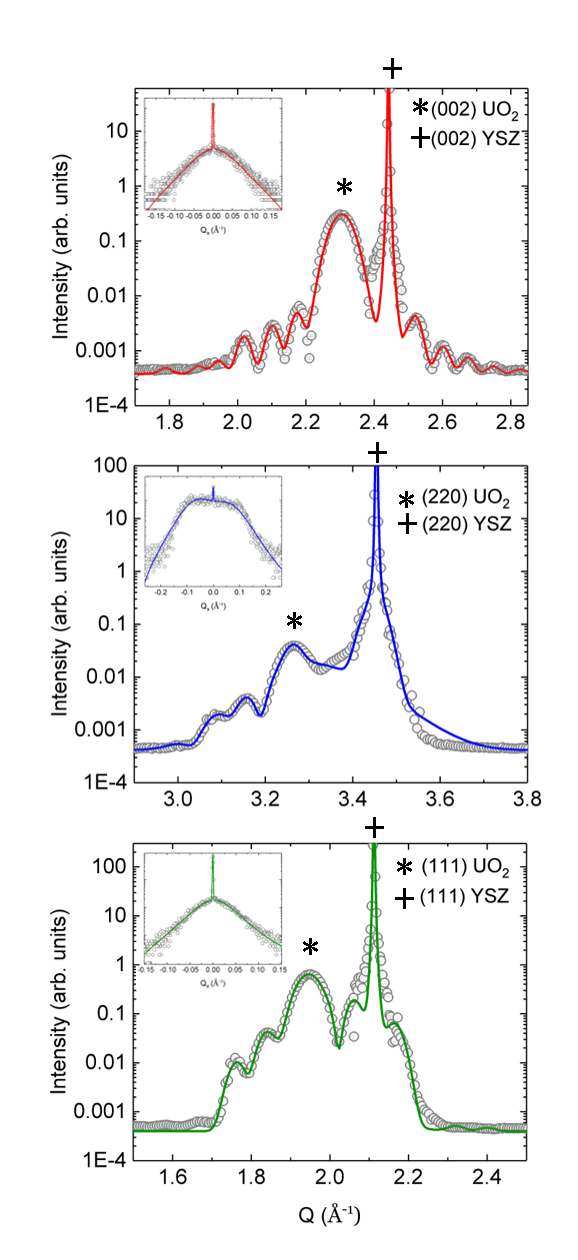}
	\caption{Longitudinal $\omega - 2\theta$ measurements taken of the UO$_2$ (002), (220) and (111) Bragg peak for the (001), (110) and (111) oriented films respectively, where the data are shown as open grey circles, and the corresponding fits as solid  red, blue and green lines. The insets show the corresponding rocking curve measurements taken at each Bragg peak. Rocking curves have been fitted with a sharp Gaussian peak and broader Lorentzian squared component, the fit envelope is shown by the solid red, blue and green lines, for the (001), (110) and (111) oriented films respectively.}
	\label{fig:ctr_model}
\end{figure}

\begin{table}[h!]
\centering 
{\renewcommand{\arraystretch}{1.5}
\begin{tabular}{c c c c c} 
\hline \hline 
    Sample& t$_{UO_2}$ (\AA) & $\sigma_{UO_2}$ (\AA) & t$_{UO_X}$ (\AA) & $\sigma_{UO_X}$ (\AA) \\ [0.5ex] 
\hline 
(001) UO$_2$ & 68.8(3) & 7(1)& 11.4(8) & 4(1)  \\ 
(110) UO$_2$ & 71.0(1)& 8(1)& 15.0(3)& 6(2)   \\ 
(111) UO$_2$ & 69.0(6)& 3.4(1)& 7.0(1)& 4(1)   \\ 

\hline 
\end{tabular}}
\caption{Summary of the optimised parameters, obtained from the fits of the XRR data shown in Fig \ref{fig:pristinexrr111}, where t$_{UO_2}$ and $\sigma_{UO_2}$ are the thickness and the root mean squared roughness of the UO$_2$ layer, and t$_{UO_X}$ and  $\sigma_{UO_X}$ are the thicknesses and roughnesses of the $UO_X$ layer.} 
\label{tab:reffitpars1}
\end{table}

Figure \ref{fig:pristinexrr111} shows the XRR profiles recorded for the (001), (110) and (111) UO$_2$ thin films, where the data are shown as open grey circles and the fits are displayed as solid red, blue and green lines, respectively. The corresponding inset for each reflectivity spectra displays the scattering length density (SLD) profile. As shown, sample was modelled as three distinct layers: the YSZ substrate (grey), UO$_2$ film (dark yellow), and a higher oxide UO$_X$ phase (light yellow), with the UO$_X$ layer consisting of a series of UO$_2$ monolayer slices. Each layer is defined by the material density  (r$_{sub}$, r$_{UO_2}$, r$_{UO_X}$) given in molecules per\,\AA$^3$, the layer thickness (t$_{sub}$, t$_{UO_2}$, t$_{UO_X}$) in $\AA$ngstroms, and the roughness of each interface ($\sigma$$_{sub}$, $\sigma$$_{UO_2}$, $\sigma$$_{UO_X}$), measured as the root mean squared of the fluctuations in the height of the layer (\AA).

The layer densities of the substrate and stoichiometric UO$_2$ layer were fixed at theoretical values of  0.0266 molecules per\,\AA$^3$ and 0.02445 molecules per\,\AA$^3$ respectively, whereas the density of the slices comprising the UO$_X$ layer were allowed to vary, each having a lower density than the stoichiometric UO$_2$ value and decreasing in value on approach to the film surface. This process was applied for each of the three samples; he corresponding fit parameters are given in Table \ref{tab:reffitpars1}, where the quoted thickness parameter for the UO$_X$ layer assumes half the density of UO$_2$, the average between the bulk value and air, to give an average layer thickness.

In addition to XRR, the structural properties of each film were also characterised using longitudinal $\omega -2\theta$, rocking curve and off-specular XRD measurements, enabling the film epitaxy to be determined in both the specular direction and in the plane of the films. Figure \ref{fig:ctr_model} displays the longitudinal $\omega-2\theta$ spectra recorded for each sample. The data are fitted with a series of Gaussian peaks to model the Laue fringes extending out from the Bragg peak, and an asymmetrical Lorentzian squared term to reproduce the sharp substrate Bragg peak. Using these fits, the central Bragg peak positions were extracted, enabling the lattice spacing, d$_{(hkl)}$, of both the film and substrate to be determined. These values are given in Table \ref{tab:offspecpars} alongside the corresponding film strain and percentage deviation from bulk lattice spacings. Although not shown here, off-specular diffraction scans, were also performed to ensure the films were epitaxial in the plane of the film in addition to the specular direction. Although the films demonstrated in-plane epitaxy, only a limited number of off-specular peaks were explored and as such in-plane lattice parameters were not well characterised and are therefore not reported here.

The corresponding rocking curve measurement taken at the specular UO$_2$ Bragg peak for each sample is displayed in the inset of Fig. \ref{fig:ctr_model}.  For the (001) and (111) oriented films, a two component rocking curve profile is observed, where the broad component has been fitted with a Lorentzian squared term and the narrow component with a Gaussian function.

For the (110) film surface, the mosaic is different. Although the overall roughness in the specular direction is similar to the (001) surface, the off-specular intensity can no longer be fit with just a simple Lorentzian squared peak centered at Q$_x$\,=\,0. The central sharp component is much smaller for this surface than for the other two principle directions, and the scattered intensity falls off much slower. This could reflect a more complicated surface structure, that may perhaps be related to the (110) surface being the only polar UO$_2$ surface investigated.

\begin{table}[!h]
\centering 
{\renewcommand{\arraystretch}{1.5}
\begin{tabular}{c c c c c} 
\hline \hline 
    Sample & d$_{UO2}$ (\AA) & $\%$ bulk dev. & d$_{YSZ}$ (\AA) & $\%$ bulk dev. \\ [0.5ex] 
\hline 
(001) UO$_2$ & 5.47(2) & -0.4   & 5.15(2) & $+0.7$ \\ 
(110) UO$_2$ & 3.82(2) & - 1.2  & 3.45(2) & +4.8   \\ 
(111) UO$_2$ & 3.22(2) & +2.0 & 2.97(1)  &  +0.4 \\ 
\hline 
\end{tabular}}
\caption{The specular lattice spacings calculated for each film orientation when compared with bulk UO$_2$ lattice spacing values.} 
\label{tab:offspecpars}
\end{table}

\subsection{UO$_2$ Dissolution}

Prior to exposing the sample to the x-ray beam, XRR measurements were initially recorded both before and after exposing the sample to water. No change in the reflectivity profiles were observed, demonstrating that any changes in the surface morphology is a response of radiolytic dissolution through exposure of the water to the synchrotron beam.

To induce radiolysis of the water, the thin film surface tension cell was exposed to the synchrotron x-ray beam for a selected exposure time. During this exposure, it is necessary to fix a low incident beam angle, such that the footprint of the beam extends across the entire sample in the direction parallel to the beam. Given the beam was focussed to 100\,$\mu$m  in the vertical direction, the incident beam angle $\omega$, was fixed at 0.5 $^{\circ}$ such that the footprint extended to 11.5\,mm, entirely covering the length of the film. To ensure the volume of water between the Kapton film and the sample surface remained constant during all measurements, a series of procedures were followed. Firstly, on aligning each sample the beam was positioned centrally along both x and y dimensions of the sample. Once aligned, the Kapton dome was constructed parallel to the x-ray beam as shown in Fig. \ref{fig:thinfilmcell}. This ensured that the volume of water above the sample surface varied only in the direction perpendicular to the beam and not parallel, such that the volume remains constant at all points along the beam footprint. Additionally, during repeat experiments of increasing exposure times, the Kapton dome was identically reconstructed using fixed markers placed on the stainless-steel stub.

\begin{figure*}
	\centering
	\includegraphics[height=0.85\textheight]{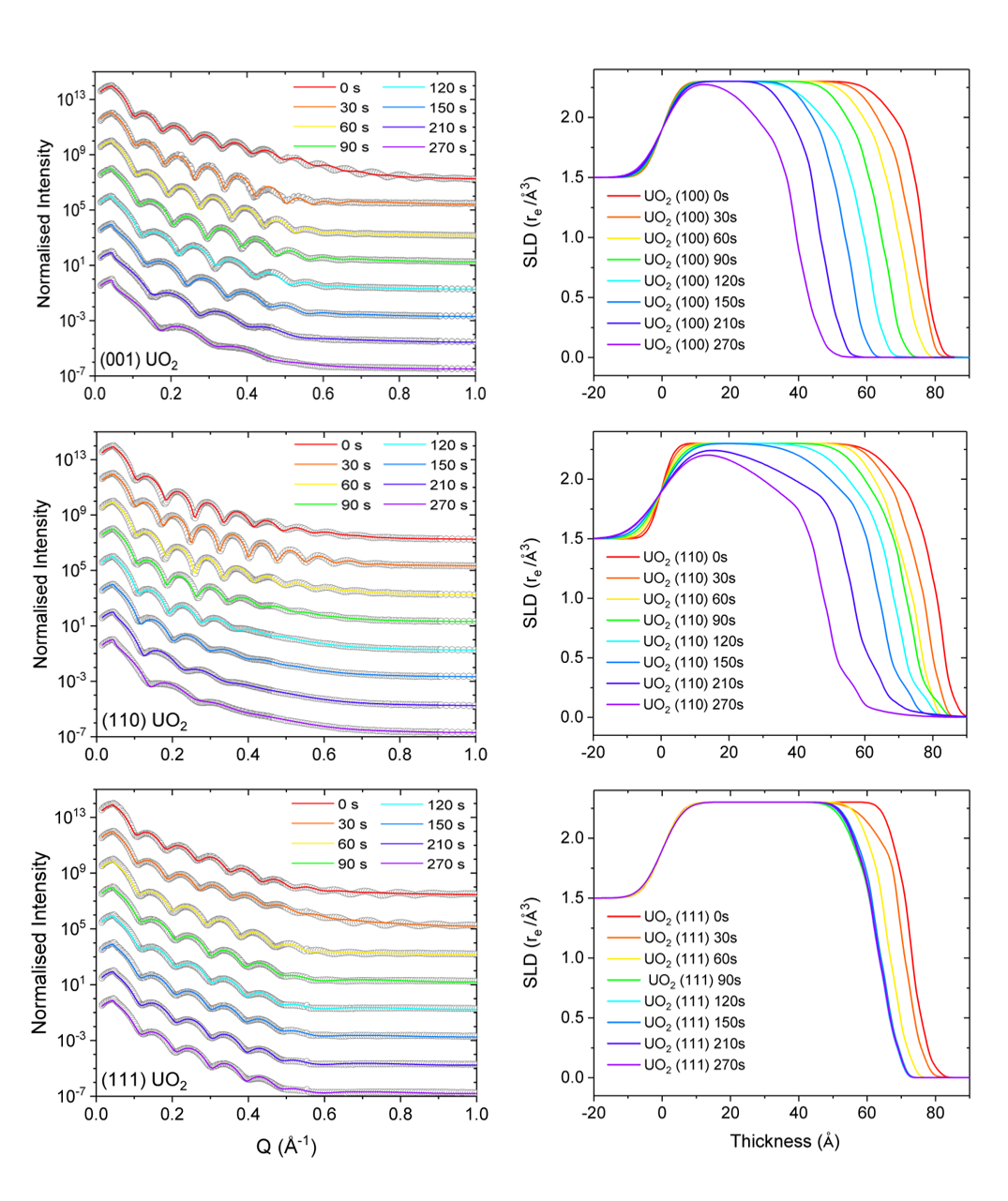}
	\caption{Panel (a) shows X-ray reflectivity data and panel measured at exposure times of 0 s, 30 s, 60 s, 90 s, 120 s, 150 s, 210 s and 270 s for each sample; the experimental data are represented by the open grey circles and the fitted calculations by the solid red, orange, yellow, green, light blue, blue, dark blue, and purple lines, respectively. Each XRR profile has been normalised to 1, and offset for comparison. Panel (b) shows the corresponding scattering length density (SLD) plots as a function of sample depth,  where 0 \AA\, represents the substrate / UO$_2$ interface.}
	\label{fig:corrodedxrr2}
\end{figure*}

The effect of crystal orientation on the dissolution of epitaxial UO$_2$ thin films has been investigated for three principle orientations: (001), (110), and (111). Figure \ref{fig:corrodedxrr2} shows the x-ray reflectivity spectra (left) and associated SLD plots (right) for each film, after corrosion times of 0s, 30s, 60s, 90s, 120s, 150s, 210s and 270s. The data are shown as open grey circles, and the fitted calculations are given by the solid red, orange, yellow, green, light blue, blue, dark blue, and purple lines,for increasing corrosion times. All XRR profiles (Fig. \ref{fig:corrodedxrr2}(left)) are shown to vary as a function of corrosion, with the Kiessig fringes both increasing in separation and becoming further dampened, indicating a reducing film thickness and increasing film roughness as a function of corrosion time. The corresponding SLD profiles (Fig. \ref{fig:corrodedxrr2}(right)) depict this behaviour more clearly, with the total film thickness reducing over the course of the corrosion, and the electron density profile becoming more graduated between the UO$_2$ and UO$_X$ boundary.

\section{DISCUSSION}
\subsection{Effect of Crystallographic Orientation }
Characterisation of the (001), (110) and (111) UO$_2$ films by x-ray reflectivity prior to the dissolution experiments, showed the films to have similar UO$_2$ starting thicknesses of approximately 70 \AA. Each sample was additionally found to comprise an oxidised surface layer, UO$_X$, however, a variation in thickness of this layer was observed across the different orientations. As detailed in Table \ref{tab:reffitpars1}, the thickness varied from 7 \AA\, to 15 \AA,\, with the thinnest UO$_X$ layer found on the (111) surface.

The observed differences in starting UO$_X$ thickness (Table \ref{tab:reffitpars1}) may be attributed to the surface terminations of each film orientation.
As the (110) oriented film is the only polar, stoichiometric UO$_2$ surface investigated, we suggest that the enhanced surface oxidation is caused by the increased exposure of uranium atoms at the film/air interface. This is in contrast to the non-polar, oxygen terminated (111) and (001) surfaces, where the uranium atoms are buried. The (111) surface is however known to be significantly more stable than the (001) \cite{Bottin2016}, which in fact is not predicted to be stable by theoretical calculations. The higher surface energy of the (001) surface may therefore account for the increased UO$_X$ thickness observed for the (001) film compared with the more stable (111).

X-ray diffraction measurements of the specular reflections for each film, enabled the specular lattice spacings, d$_{(hkl)}$, to be determined, from which the percentage deviation from the bulk lattice parameter was calculated, Table II. These induced strains showed slight variations across the different orientations, however, the differences are not believed to have significant effect on the corrosion rates.

To induce radiolytic dissolution of the UO$_2$ films, each sample was exposed to MilliQ water in the presence of the x-ray beam for a series of exposure times, after which XRR profiles were collected. For the (001) and (110) oriented films, the separation of the reflectivity fringes is shown to significantly increase over the 270\,s corrosion timescale, indicating a decrease in the film thickness as demonstrated by the SLD profiles in Fig. \ref{fig:corrodedxrr2}\,(right) However, the profiles displayed for the (111) film, show a less significant change over the same exposure times, suggesting that the (111) orientation of UO$_2$ is less susceptible to the radiolytic environment.
\begin{figure*}

	\centering
	\includegraphics[height=0.3\textheight]{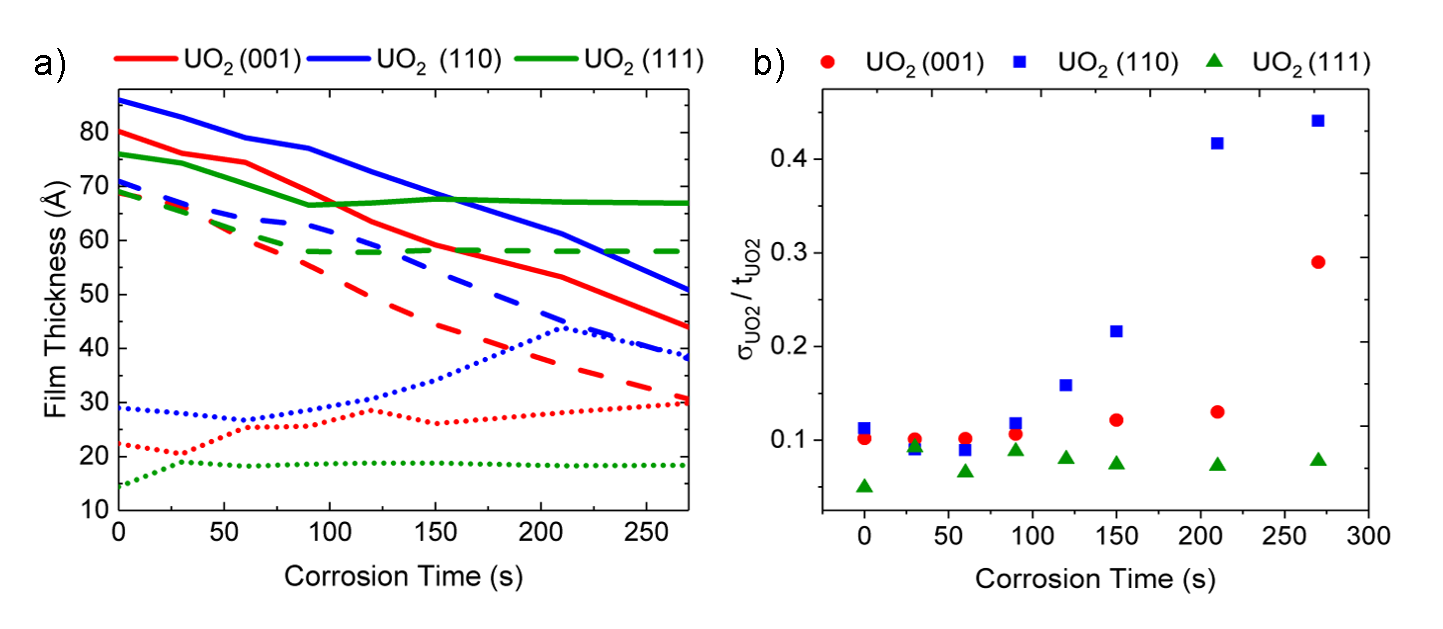}
	\caption{(a) The layer thicknesses for the (001), (110) and (111) UO$_2$ thin films, shown in red, blue and green respectively, are compared as a function of corrosion time. The UO$_2$ and total film thickness are displayed by the dashed and solid line, respectively. The dots indicate the dissolution zone – see text for definition. The error bars here are about +/- 3 \AA. (b) The roughness of the UO$_2$ layer normalized by the total UO$_2$ thickness for the 3 different directions. Errors are $\sim$ 0.05.}
	\label{fig:corrodedxrr5}
\end{figure*}

\begin{figure*}
	\centering
	\includegraphics[height=0.23\textheight]{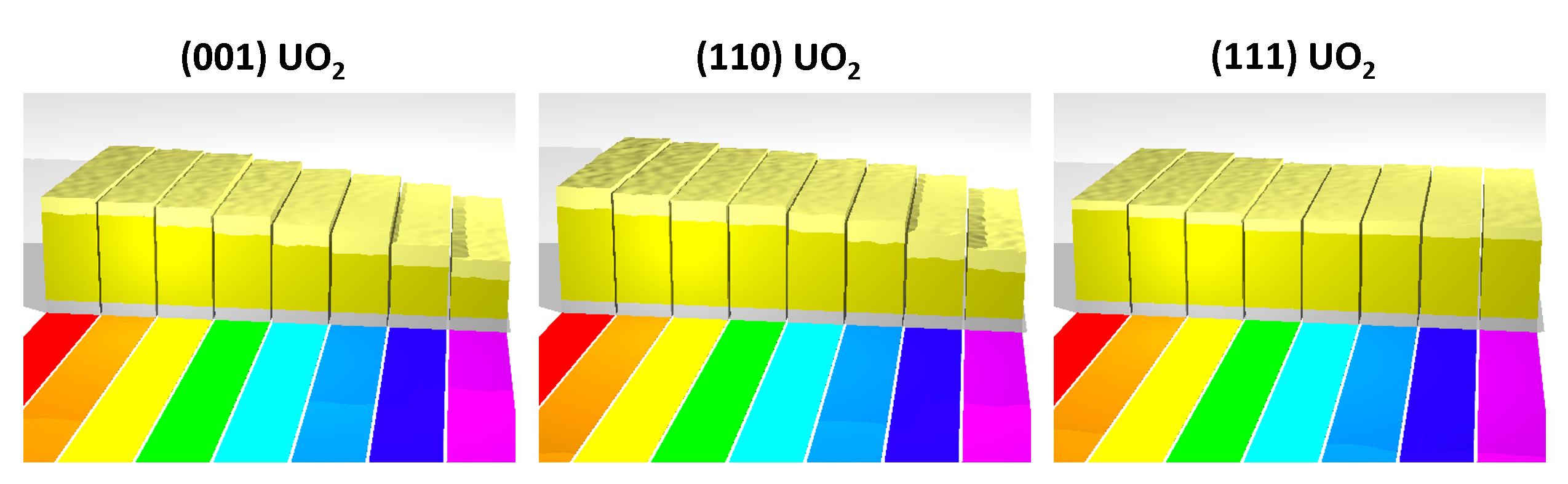}
	\caption{Schematic diagrams of the UO$_2$ (001), (110) and (111) thin films as a function of dissolution. Dissolution is shown to progress from left to right in each figure, representing the measured time steps of 0\,s, 30\,s, 60\,s, 90\,s, 120\,s 150\,s, 210\,s and 270\,s. The substrate is represented in grey, with the UO$_2$ and UO$_X$ layers shown in dark and light yellow respectively. Figure produced using POV-Ray.}
	\label{fig:POV}
\end{figure*}

Through analysing the XRR spectra after each recorded corrosion interval, the thickness parameters of both the UO$_2$ and UO$_X$ layers were extracted in order to compare the rate of dissolution for each film orientation. Figure \ref{fig:corrodedxrr5}(a) displays the layer thicknesses comprising the (001), (110), and (111) oriented thin film samples as a function of corrosion time, represented by the red, blue, and green lines. The UO$_2$ and UO$_X$ thicknesses are shown as dashed and solid lines, respectively. Shown as dotted points are the change of the `dissolution thickness', which we define as the sum of the UO$_2$ roughness, plus the thickness and roughness of the UO$_X$ layer as a function of exposure. As shown, between 0 and 90\,s, each UO$_2$ orientation is observed to corrode at an approximately comparable rate, however after 90\,s a distinct difference is observed in the corrosion profiles. While the (001) and (110) orientations continue to corrode at an approximately linear rate, further corrosion of the (111) oriented film is not observed.

For the (111) oriented UO$_2$ film, a total of 9.6\,\AA\, of material is found to have been removed from surface after 90\,s of corrosion. This thickness approximately correlates with the loss of one (111)-UO$_2$ monolayer. This suggests that after the removal of the surface layer, which may be more susceptible to dissolution as a result of surface defects or oxidation, the (111) UO$_2$ thin film is prevented from further corrosion as a result of surface termination.

Additionally, Fig. \ref{fig:corrodedxrr5}(b) highlights the change in roughness, by plotting the roughness of the UO$_2$ layer, normalised by the thickness of the remaining UO$_2$ layer.


Combining the extracted layer thickness and roughness values extracted from the XRR analysis, it is possible to build up a model of the films morphology as a function of corrosion time, this is shown schematically in Fig. \ref{fig:POV}, where the dissolution is shown to progress from left to right, and the substrate, UO$_2$ and UO$_X$ layers are again displayed in grey, dark yellow and light yellow respectively.

As the UO$_2$ (111) surface is well known to be entirely oxygen terminated \cite{Bottin2016,Ellis1980}, and the most stable of the principal crystallographic orientations, it is perhaps unsurprising that the (111) UO$_2$ film is less susceptible to corrosion in comparison with the (110) and (001) orientations. However, although the (111) UO$_2$ surface may be anticipated to undergo dissolution at a slower rate than the (001) and (110) surfaces, the observation that the corrosion is completely halted after the loss of the initial surface layer is significant. This result may have important consequences for theoretical dissolution models, as it is evident that orientation dependence must be taken into consideration when modelling rate of dissolution of UO$_2$ surfaces.

It is clear that a UO$_{2+x}$ layer is present on all surfaces prior to dissolution, however determining how this layer changes as a function of corrosion in challenging. As Fig. \ref{fig:corrodedxrr5}(a) shows, the thickness of the dissolution zone increases for the (001) and (110) surfaces. The density of the top layers are certainly below that of UO$_{2.00}$ (see Fig. \ref{fig:corrodedxrr2}), however XRR cannot determine the exact stoichiometry of the UO$_{2+x}$ layer. XPS measurements have been attempted, but the changes in the spectra are small, leaving an uncertainty in this approach \cite{Rennie2018}. One problem is the lack of reliable standards for XPS spectra in the different valence states in the uranium oxides.

\subsection{X-ray induced radiolysis}
Within this study x-rays were utilised to induce water radiolysis, however it is noted that the vast majority of UO$_2$ dissolution studies are primarily concerned with replicating radiolytic conditions induced by alpha radiation \cite{Shoesmith2000,Sunder2004,Sunder1997,Bailey1985,Wren2005}. This is because these studies focus on the radiolytic environments predicted to occur thousands of years into long term spent nuclear fuel storage, i.e. when the $\beta$ and $\gamma$ fields have significantly decayed. Given that there is potential for UO$_2$ fuel to come into contact with water earlier in the fuel cycle, it is important to additionally assess the effect of gamma radiation, which x-ray radiation more accurately represents.

\begin{table*}
\centering 
{\renewcommand{\arraystretch}{1.5}
\begin{tabular}{c c c c c c c c} 
\hline\hline 
Radiation & G(-H$_2$O) & G(H$_2$) & G(H$_2$O$_2$) & G(e$_{aq}^-$)& G(H$^\bullet$)& G(OH$^\bullet$)& G(HO$_2^\bullet$)\\ [0.5ex] 
\hline 
$\gamma$, fast $\beta$ & -0.43 & 0.047 & 0.073 & 0.28 & 0.062 & 0.28 & 0.0027 \\ 
12MeV $\alpha$ & -0.294 & 0.115 & 0.112 & 0.0044 & 0.028 & 0.056 & 0.007 \\ [1ex]

\hline 
\end{tabular}}
\caption[A comparison of product yields (G values) of radiolytic species, generated for $\gamma / fast \beta$ and $\alpha$ irradiation of neutral water.]{A comparison of product yields (G values) of radiolytic species, generated for $\gamma / fast \beta$ and $\alpha$ irradiation of neutral water \cite{Choppin1995}.} 
\label{table:g_values2}
\end{table*}

In particular, it is important for the effect of alternate types of radiation to be studied in isolation, in order to determine whether different ionising species affect the rate and mechanism of UO$_2$ dissolution. Table \ref{table:g_values2} highlights significant differences in radiolytic yields, known as G values determined by Choppin \textit{et al.} \cite{Choppin1995}, for both $\gamma$ radiation and 12 MeV $\alpha$ particles. As shown, for $\alpha$ radiation, H$_2$O$_2$ is the most dominant radiolytic species, whereas for $\gamma$ radiation OH$^{\bullet}$ becomes more important. Although the different oxidants all act to drive UO$_2$ dissolution, the mechanism for oxidation and thus the rate of dissolution, is dependent upon the radiolytic species.

A significant body of work has been conducted to investigate UO$_2$ dissolution under a range of radiolytic conditions, however the role of individual oxidants is not well understood \cite{Ekeroth2006}. Studies performed by Ekeroth \textit{et. al} \cite{Ekeroth2006}, show that despite the increased yield of OH$^{\bullet}$ and HO$^{\bullet}_2$ for $\gamma$ radiolysis, the short lifetimes of these species result in significantly higher steady state concentrations of H$_2$O$_2$ and O$_2$. Even considering the increased oxidation rate of UO$_2$ for the radical species, it is predicted that under gamma radiation H$_2$O$_2$ and O$_2$ are the key drivers of UO$_2$ dissolution \cite{Ekeroth2006}.

However, as discussed in our previous publication \cite{Springell2015}, the corrosion observed within our x-ray induced dissolution experiments is distinctly confined within the path of the x-ray beam. This may suggest that the shorter lived radical species, which are confined within the beam path, have a much more significant effect on the dissolution mechanism.

To further investigate the role of short lived species in greater detail, it is necessary to isolate the effect of individual corrosion species, however, conducting such experiments is increasingly complex. Information on the relative concentrations of oxidant species could be obtained through compositional analysis of the water, however given the short lifetimes of some of the generated oxidants, \textit{in-situ} analysis would be required. While this in itself presents a significant experimental challenge, the additional limitation of having a very small area of water irradiated by the x-ray beam makes performing such measurements on the I07 beamline impossible. An alternative experimental method would be to introduce a chemical species to the water solution such that specific oxidant species are scavenged and thus do not contribute to the dissolution. However, introducing additional chemical species to scavenge alternate oxidant products would significantly alter the chemistry of the water solution, introducing an additional experimental parameter that would make the results more difficult to interpret.

An alternative approach would be to conduct a comparative set of dissolution experiments whereby water radiolysis is induced using alpha radiation. In this way, corrosion rates can be generated for two distinct water chemistries, comprising vastly different oxidant compositions, and thus producing a set of simultaneous rate equations allowing the influence of individual parameters to be determined. With the advent of high energy irradiation beamlines, high fluxes  of alpha particles can be produced so that such experiments could be conducted.

\section{CONCLUSIONS}
The effect of crystallographic orientation on the radiolytic dissolution rate of UO$_2$ has been investigated using (001), (110) and (111) oriented UO$_2$ thin films exposed to water in the presence of a high flux synchrotron x-ray beam. Analysis of x-ray reflectivity measurements recorded over a total corrosion time of 270s enabled the film morphologies to be determined as a function of exposure time. Initially, corrosion is observed to progress at comparable rate for each crystallographic orientation, however after 90\,s of exposure, no further corrosion is observed for the (111) orientated UO$_2$ thin film. This contrasts with the (001) and (110) oriented films that are observed to continually corrode at an approximately linear rate. The stark contrast in dissolution behaviour of the (111) oriented film is attributed to the surface termination. This result may have important consequences for theoretical dissolution models, as it is evident that orientation dependence must be taken into consideration to obtain accurate long-term predictions of the dissolution behaviour of UO$_2$.
While it has been demonstrated that both the x-rays and the water interface are essential to induce corrosion, obtaining  a full understanding of the mechanisms responsible for UO$_2$ dissolution requires further investigation. This topic remains a broad field of research, and the results presented here highlight the need to understand the role of individual oxidant species on the corrosion mechanism.

\section{REFERENCES}

\bibliography{bibliog4}{}
\end{document}